\documentclass[aps,prb,preprint,letterpaper,floatfix,showpacs]{revtex4}

\usepackage[scriptsize,normal]{subfigure}
\usepackage{graphicx}
\usepackage{amsmath}
\makeatletter

\usepackage{amsfonts}

\newcommand{\vs}{\mbox{\large \boldmath $\sigma$}}

\newcommand{\tr}{\mathrm{tr}}

\newcommand{\1}{1\negthickspace{\rm I}}

\newcommand{\ie}{\mbox{i.\ e.\ }}
\newcommand{\eg}{\mbox{e.\ g.\ }}

\newcommand{\etal}{\mbox{\it et.\ al.\ }}

\newcommand{\PRB}{\mbox{Phys.\ Rev.\ B}}
\newcommand{\PRL}{\mbox{Phys.\ Rev.\ Lett.}}
\newcommand{\NLSM}{\mbox{NL$\sigma$M}}

\newcommand{\NMR}{kumagai97,magishi98,carretta98}
\newcommand{\INS}{eccleston98,katano99}
\newcommand{\RAMAN}{gozar}
\newcommand{\allexp}{\NMR,\INS,\RAMAN}
\newcommand{\ARPES}{takahashi}

\begin{document}
\bibliographystyle{apsrev}

\title{Doping dependence of the spin gap in a 2-leg ladder}

\author{L. \surname{Campos Venuti} and A. Muramatsu}

%\email{campos@theo3.physik.uni-stuttgart.de}

\affiliation{Institut f{\"u}r Theoretische Physik III, Universit{\"a}t
Stuttgart, Pfaffenwaldring 57, D-70550 Stuttgart, Federal Republic of Germany}

\begin{abstract}
A spin-fermion model relevant for the description of
cuprates ladders is studied in a path integral formalism,
where, after integrating out the fermions, an effective action for the spins 
in term of a Fermi-determinant results. The determinant can be evaluated 
in the long-wavelength, low-frequency limit
to all orders in the coupling constant, leading to a non-linear 
$ \sigma  $ model with doping dependent coupling constants. An explicit evaluation
shows, that the spin-gap diminishes upon doping as opposed to previous 
mean-field treatments.
\end{abstract}

\pacs{71.10.Fd,71.27+a,75.10.Jm}

\maketitle

\section{Introduction}

Doped quantum antiferromagnets (QAFM) constitute a major unresolved problem 
in condensed matter physics, which is at the center of current research
since the discovery of high $ T_{c} $ superconductivity
\cite{bed-mull}. In particular, the case of a doped spin
liquid --where no symmetry is spontaneously broken-- is very
challenging, since the starting 
point, the spin liquid state, cannot be described by a classical 
N{\'e}el state.

This problem is not only of theoretical relevance. 
$ \textrm{Cu}_{2}\textrm{O}_{3}$ ladders are present in $
\textrm{Sr}_{14-x}\textrm{Ca}_{x}\textrm{Cu}_{24}\textrm{O}_{41} $
and many experiments support the presence of a spin gap and a finite
correlation length
%\cite{kumagai97,magishi98,eccleston98,katano99},
\cite{\allexp}, two crucial
ingredients signaling a spin liquid state. 
With isovalent $ \textrm{Ca}^{2+} $ substitution of $ \textrm{Sr}^{2+} $
holes are transferred from the $ \textrm{CuO}_{3} $ chains to the
ladders~\cite{osafune97},
increasing the conductivity of the latter. The spin gap, as measured by Knight
shift or NMR experiments \cite{\NMR}
is seen to diminish. With increasing doping,
superconductivity 
is ultimately stabilized under pressure
\cite{supercond-sci,supercond2}, a 
phenomenon that suffices to justify the interest for the subject.

The simplest model which is believed to grasp the physics of the problem is
the $ t-J $ model on a two leg ladder. 
It is believed in general that this system evolves continuously 
from the isotropic case to the limit of strong rung interaction. In this 
limit some simplifying pictures are at hand: without doping the gap
is the energy of promoting a singlet rung to a triplet ($\sim J_{\perp}$).
Interaction among the rungs leads eventually to the usual magnon
band. 
Upon doping the systems shows two  
different kinds of spin excitations~\cite{two-exc1,two-exc2}.
One is still the singlet-triplet transition as before,
the other kind corresponds to the splitting of a hole
pair into a couple of 
quasiparticles (formed by a spinon and an holon), each carrying charge
$ +\left| e\right|  $ and spin 1/2. The number of possible
excitations is proportional to $ \left( 1-\delta \right)  $ (for the magnons)
and $ \delta  $ (for the quasiparticles), respectively, where $
\delta  $ is   
the number of holes 
per copper sites. For this reason, at low doping concentration,  the
magnon gap will be the most important in influencing the form of the
static susceptibility or dynamical structure factor. 

First Sigrist \etal~\cite{sigrist-mf} and more recently Lee \etal
\cite{lee-mf} attacked the problem ultimately with some sort of
mean field decoupling. Their results agree in predicting an increase
of the magnon gap ($\Delta_M$, originated from the singlet-triplet
transition), while Lee \etal were also able to calculate a 
decrease of the quasiparticle gap ($\Delta_{QP}$ originated from the
splitting of a hole pair) for small doping
concentrations. 

In contrast to the mean-field results above,  Ammon \etal~\cite{ammon-TDMRG}
obtained a decrease of the magnon gap and an almost doping independent
$ \Delta _{QP} $ using temperature density matrix renormalization group 
(TDMRG). As already mentioned,
a  decrease of the spin gap is also observed in a number of experiments
\cite{kumagai97,magishi98,carretta98}. 

In this paper we concentrate on the behavior of the magnon gap upon doping.
Due to the contradiction above it is imperative to go beyond mean field
and include the role of fluctuations in a controlled manner. 
A mapping from an AFM Heisenberg model to an effective
field theory, the non linear $\sigma$ model (\NLSM), proved very efficient in
describing the magnetic properties of two dimensional spin lattices
\cite{CHN}, chains~\cite{haldane1}, and 
ladders~\cite{morandi}. This mapping was extended in Ref.\
\cite{mu-zey} to the case   
of a doped two dimensional QAFM using a procedure that we will closely follow.

\section{Mapping to an effective spin action}

Since no satisfactory analytical treatment of
the $t-J$ model away from half filling is possible at present, we
focus on the so called spin-fermion model. 
This Hamiltonian can be derived in fourth order degenerate
perturbation theory~\cite{zaanen-pd-sf,mu-pd-sf} from the $ p-d $, three band, 
Emery model~\cite{emery-pd}, that gives a detailed description of the 
cuprate materials. There the role of perturbation is played by the
hybridization  
term between the $ p $-orbital (oxygen) and the $ d $-orbital 
(copper). A further simplification of the
model was proposed by Zhang and Rice~\cite{ZR-pd-tj}, that leads to the
$ t-J $ model. 

\begin{figure}
\begin{center}
%\psfrag{JPAR}[b]{$J_{||}$}
%\psfrag{JPER}[b]{$J_{\perp}$}
\includegraphics[width=6cm,height=3.5cm]{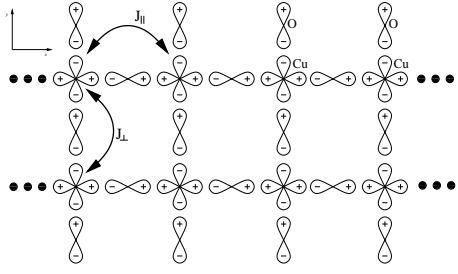}
\end{center}
\caption{\footnotesize Schematic picture of a two leg ladder Copper-Oxide.}
\label{laddfig}
\end{figure}

A typical copper-oxide two leg ladder, as those present in
$\textrm{Sr}_{14-x}\textrm{Ca}_{x}\textrm{Cu}_{24}\textrm{O}_{41} $
is depicted in Fig.\ \ref{laddfig}. It is generally accepted that the dopant
holes reside on \emph{p}-orbitals
on the oxygens sites, whereas on the $ \textrm{Cu}^{2+} $ ions a localized
hole resides, represented by
spin 1/2 operator which interact via a nearest neighbor exchange.

The spin-fermion Hamiltonian is defined as follows:

\begin{eqnarray}
H_{SF}&=&t\sum_{\left\langle j,j';i\right\rangle ,i,\sigma }\left( -1\right)
^{\alpha_{i,j}+\alpha_{i,j'}} c^{{\dag} }_{j,\sigma } c_{j',\sigma} +J_{K}\sum_{i} 
\mathbf{R}_{i}\cdot \mathbf{S}_{i} \nonumber \\
&&+J_{H}\sum_{\left\langle i,i'\right\rangle } 
\mathbf{S}_{i}\cdot\mathbf{S}_{i'} . \label{first} 
\end{eqnarray}

The index $i$ ($j$) runs over the Cu (O) sites, $c^{{\dag} }_{j,\sigma }$
creates a hole in an oxygen $p$ band and $\mathbf{S}_{j}$ are spin
operators for the copper ions. The coefficients $ \alpha_{i,j} $ take
care of the 
sign of the p-d overlap and $ \alpha_{i,j}=1 $ if
$ j=i+\frac{1}{2}\hat{x} $ or $ i+\frac{1}{2}\hat{y} $ and $
\alpha _{i,j}=2 $ if $ j=i-\frac{1}{2}\hat{x} $ or $
i-\frac{1}{2}\hat{y} $. Finally the operator $ \mathbf{R}_{i} $ is defined as

\begin{equation}
\mathbf{R}_{i}=\sum_{\left\langle j,k;i\right\rangle ,i,\alpha ,\beta}
( -1)^{\alpha_{i,j}+\alpha_{i,k}}c^{{\dag} }_{j,\alpha}
\vs_{\alpha ,\beta }c_{k,\beta } .
\end{equation} 

Following~\cite{ZR-pd-tj} we can define the following operator centered on
the copper site $ P_{i,\sigma }=\left( 1/2\right) \sum _{\left\langle j;i\right\rangle }\left( -1\right) ^{\alpha _{i,j}}c_{j,\sigma } $
which represents non orthogonal orbitals with a high weight on the $
i $ site. 
Their anti-commutation relations are
\begin{equation}
\left\{ P_{i,\sigma },P^{{\dag} }_{i',\sigma '}\right\} =\delta_{\sigma ,\sigma '}\left( \delta_{i,i'} -\frac{1}{4}\delta_{\left\langle i,i'\right\rangle }\right),
\end{equation} 
and we can rewrite the Hamiltonian in terms of these operators as follows

\begin{eqnarray}
H_{SF} & = & 4t\sum_{\substack{l=1\ldots L \\ \lambda =1,2,\sigma }} P_{l,\lambda,\sigma  }^{{\dag} }P_{l,\lambda,\sigma}
+4J_{K}\sum_{\substack{l=1\ldots L\\ \lambda =1,2,\alpha,\beta}}P_{l,\lambda,\alpha}^{{\dag} }\vs_{\alpha,\beta} P_{l,\lambda,\beta}\cdot
\mathbf{S}_{l,\lambda }
\nonumber \\ & &
+J_{\perp }\sum_{l=1\ldots L}\mathbf{S}_{l,1}\cdot
\mathbf{S}_{l,2}+J_{||}\sum_{\substack{l=1\ldots L\\\lambda =1,2}}\mathbf{S}_{l,\lambda }\cdot
\mathbf{S}_{l+1,\lambda } ,\label{second}
\end{eqnarray} 
$L$ is the number of the rungs along the ladder and $\lambda =1,2$
distinguishes the two  legs.
For the sake of generality, an anisotropy in the
Heisenberg term is allowed. 

%Written in this way it is now evident that we have reduced to one the
%number of fermionic states per copper site, whereas originally we had
%five oxygens every two coppers. Some may argue that such a reduction
%would be impossible in presence of a direct oxygen-oxygen nearest
%neighbor $t'$. We believe our result (\eg basically a decrease of the
%magnon gap with doping) remain correct for more complicated band
%structure. On the other hand since we want to compare our theory with
%experiments it is preferred to keep the number of parameters as low as
%possible.
%
%One can see that in Eq.~(\ref{first}) we did not include any direct
%oxygen-oxygen hopping term which is sometimes included in spin fermion
%models. This enabled us to lower the number of fermionic states
%per rung from 10 to 4 and to come to  Eq.~(\ref{second}).
%Since we wanted to compare our results with available experiments, our
%choice was dictated by the intention of keeping the number of 
%parameter as low as possible.

The different steps of our procedure are the following: first find
orthogonal (Wannier states) 
for the holes, then go to a (coherent states) path integral formulation for
spins and fermions and perform the Gaussian integration of the
fermionic degrees of freedom. The remaining part of the calculation is
devoted to the evaluation 
of the resulting Fermi determinant in the long-wavelength
low-frequency limit. 
This expansion includes the coupling constant $ J_{K} $ to all order.

Wannier states are easily find via $ P_{\mathbf{k},\sigma
}=\sqrt{\epsilon(\mathbf{k}) }f_{\mathbf{k},\sigma} $where  
$ \epsilon \left( \mathbf{k}\right) =\left( 1-\frac{\cos \left(
k_{x} a\right) +\cos \left( k_{y}a\right) /2}{2}\right)  $.
Here $a$ is the lattice constant and we used a two dimensional Fourier transform 
where $k_{y}$ takes 
only values $ 0 $ and $ \pi/a  $ distinguishing between symmetric
(bonding) and antisymmetric (antibonding) states. The partition
function can be expressed as a path integral

\begin{equation}
Z=\int D\left[ f^{\ast }\right] D\left[ f\right] D [ \hat{\Omega} ] 
e^{-S_{SF} },
\end{equation}
where $S_{SF}=S_{h}+S_{s}$.
The action $S_{s}$ contains all terms with
spins degree of freedoms only~\cite{dombre-read}:

\begin{equation}
S_{s}=\int^{\beta }_{0}d\tau \left[ -iS\sum _{l,\lambda
}\mathbf{A}\left( \hat{\Omega }_{l,\lambda 
}\right) \cdot \frac{\partial \hat{\Omega }_{l,\lambda }}{\partial
\tau }+H_{\textrm{Heis}}\left( 
S\hat{\Omega }\left( \tau \right) \right) \right] , \label{eq:s-spin}
\end{equation} 
where $\hat{\Omega }$ is a unimodular field, $S$ is the spin per site
($1/2$ in our case) and 
 $ \mathbf{A} $ is the vector potential for a (Dirac) monopole: 
$ \epsilon ^{abc} (\partial A_{a}/\partial \hat{\Omega}_{b})
=\hat{\Omega }_{c} $.  

It is by now well accepted that the effective low energy field theory
of the $d$-dimensional Heisenberg antiferromagnetic 
model is given by the 
$(d+1)$ \NLSM~\cite{haldane1,affleck,auerbach-book}.
In the case of a ladder one obtains the $(1+1)$
\NLSM~\cite{morandi,sierra}. For this reason, here we will  
deal mainly with the part of the action which contains fermionic degrees
of freedom $ S_{h} $:

\begin{equation}
S_{h}=\sum _{kq\alpha \beta }f_{k,\alpha }^{\ast }\left[ \left( i\omega
_{n}+4t\epsilon \left( \mathbf{k}\right) -\mu \right) \delta _{k,q}\delta _{\alpha
,\beta }+g\sqrt{\epsilon \left( \mathbf{k}\right) \epsilon \left( \mathbf{q}\right) }\vs
_{\alpha ,\beta }\cdot \hat{\Omega }_{k-q}\right] f_{q,\beta},\label{Sh}
\end{equation} 
here $ k=\left( k_{x},k_{y},\omega_{n}\right)  $ where $ \omega_{n}=\pi
\left( 2n+1\right) /\beta  $  
are the fermionic Matsubara frequency and $ g=4J_{K}S $. It is natural to
decompose the inverse propagator into $ G^{-1}=G_{0}^{-1}-\Sigma  $ where
the free part is 
\begin{equation}
 G_{0}^{-1}=\left( i\omega _{n}+4t\epsilon(\mathbf{k}) -\mu \right) \delta_{k,q}\delta_{\alpha ,\beta } 
,
\end{equation}
and the fluctuating external potential is 
\begin{equation}
\Sigma=-g\sqrt{\epsilon( \mathbf{k}) \epsilon( \mathbf{q}) }
\vs_{\alpha ,\beta }\cdot \hat{\Omega }_{k-q}.
\end{equation}
 
Since, according to Eq.~(\ref{Sh}) the action $S_{SF}$ is bilinear in
the fermionic variables, we can integrate them out. This leads to 
$S_{SF}=S_s -\tr \ln G^{-1}$.
Defining the matrix 
\begin{equation}
A=\sqrt{\epsilon \left(
\mathbf{k}\right) }\delta_{k,q} \delta_{\alpha ,\beta }
\end{equation}
and a rescaled propagator $\hat{G}^{-1}$ through
\begin{equation}
\hat{G}^{-1}={A^{{\dag} }}^{-1} G^{-1}A^{-1} 
\end{equation}
we can write
\begin{equation}
\tr \ln \left(G^{-1}\right) =\tr \ln \left( AA^{{\dag} }\right) +\tr \ln \left(
\hat{G}^{-1}\right),
\end{equation}
the first term gives just a constant and we can ignore
it. Again we decompose the rescaled inverse propagator as $ 
\hat{G}^{-1}=\hat{G}^{-1}_{0}-\hat{\Sigma } $ 
which brings us to
\begin{eqnarray}
\hat{G}^{-1}_{0} & = & \left( \frac{i\omega _{n}+4t\epsilon \left(
\mathbf{k}\right) -\mu }{\epsilon \left( \mathbf{k}\right) }\right) \delta_{kq} \delta_{\alpha \beta }
\equiv g_{0}^{-1}\left( k_{x},k_{y},\omega_{n}\right) \delta_{k,q}\delta _{\alpha ,\beta } ,\\ 
\hat{\Sigma } & = & -g\Omega _{\mathbf{k}-\mathbf{q},\omega -\nu }\cdot \vs _{\alpha \beta }.
\end{eqnarray}
 
The remaining part of the calculation is devoted to the evaluation
of $ S_{h\textrm{ eff}}=-\tr \ln \left( \hat{G}^{-1}\right)  $ 
in the continuum limit.

\subsection{Parameterizations}

As we already mentioned, in the undoped regime where no holes are
present, it has proven very effective a mapping from a   
antiferromagnetic Heisenberg spin ladder to a (1+1) NL$ \sigma
$M. This mapping rely on the idea that although long range order (here
antiferromagnetic) is prohibited in one dimension,
the most important contribution to the action are given by paths in
which antiferromagnetic order survives at short distance. Accordingly
the dynamical unimodular  
field is decomposed in a N{\'e}el modulated field $ \mathbf{n} $
plus a ferromagnetic 
fluctuating contribution. A gradient expansion in the dynamical field brings
then to the (1+1) NL$ \sigma  $M. The gradient expansion is justified when
the correlation length of the spin is much larger than the lattice constant
$ a $. However the prediction of the \NLSM, \ie a finite
correlation length and a triplet of massive modes above the ground 
state~\cite{haldane1,polyak,polyak-weig,shank-read} remain valid until
$ \xi \approx 2.5a $ as numerical calculations on the isotropic Heisenberg
ladder have shown~\cite{white-ladders}.  

The basic assumption of this work is then that such a parameterization
is still meaningful as long as the spin liquid state is not destroyed by
doping, as seems to be the case in experiments, where a finite
spin-gap is also seen in the doped case~\cite{\allexp}.
Then, as e.g.\ in ref.~\cite{dombre-read}, we parameterize the spin field in
the following way  
\begin{equation}
\Omega_{i,\lambda }\left( \tau \right)   =  \left( -1\right)
^{i+\lambda }\mathbf{n}_{i,\lambda 
}\sqrt{1-\left| \frac{a\mathbf{l}_{i,\lambda }}{S}\right|
^{2}}+\frac{a\mathbf{l}_{i,\lambda }}{S} ,\label{eq:param1}
\end{equation}

$\mathbf{n}_{i,\lambda }$ and $\mathbf{l}_{i,\lambda }$ are two slowly varying,
orthogonal, vector fields describing locally antiferromagnetic and
ferromagnetic configurations, respectively. $\mathbf{n}_{i,\lambda }$ is
normalized such that  $|\mathbf{n}_{i,\lambda }|^2=1$. The lattice constant
$a$ in front of $\mathbf{l}_{i,\lambda}$ in eq. (\ref{eq:param1}) makes
explicit the fact that $\mathbf{l}_{i,\lambda}$ is proportional to a generator
of rotations of $\mathbf{n}_{i,\lambda}$, namely to a
first-order derivative of $\mathbf{n}_{i,\lambda}$.

In the particular geometry of a ladder, this decomposition give rise
to two local order parameters, $\mathbf{n}_{i,1}$ and
$\mathbf{n}_{i,2}$. However we assume that spins across the chain
are rather strongly correlated such that they will sum up to give
rise to an antiferromagnetic configuration, or subtract and give a
ferromagnetic fluctuation. A further parameterization is then
\begin{equation}
\mathbf{n}_{i,\lambda }=\mathbf{N}_{i}\sqrt{1-a^{2}\left|
\mathbf{M}_{i}\right| ^{2}}+\left( -1\right) ^{\lambda }a\mathbf{M}_{i},
\end{equation}
with $ \mathbf{N}_{i}\cdot \mathbf{M}_{i} =0$ and $|\mathbf{N}_i|^2=1$.

The next step is the gradient expansion, or equivalently, in Fourier space, an
expansion in powers of $ k $. In (1+1) dimensions the field $ \mathbf{N} $
will get no scaling dimension, whereas the fields $ \mathbf{l}$ and
$\mathbf{M} $ get scaling dimension -1. Accordingly, in the subsequent
expansion we will need 
to keep terms with up to two derivative and any power of the field
$\mathbf{N}$. Terms containing  $ \mathbf{l},\mathbf{M} $ are
marginal whenever two fields or one field and one derivative are present.
Higher order terms are irrelevant and will be
discarded. This correspond to expand all our quantities up to $
O\left( a^{2}\right)  $. 

The self energy has then the following expansion
%\begin{equation}
%\hat{\Sigma }=\Sigma _{00}+a\Sigma _{01}+a^{2}\Sigma _{02}+a\Sigma _{1}+a^{2}\Sigma _{2}+O\left(
%a^{3}\right)
%\end{equation} 
\begin{equation}
\hat{\Sigma }=\Sigma _{00}+\Sigma _{01}+\Sigma _{02}+\Sigma _{1}+\Sigma _{2}+O\left(
a^{3}\right) ,
\end{equation} 
where the various quantity are 
\begin{eqnarray}
\Sigma_{00} & = & -g\delta _{k_{y}-q_{y},\pi }\, \mathbf{N}_{k_{x}-q_{x}+\pi ,\omega -\nu}\cdot
 \vs _{\alpha \beta }, \\ 
\Sigma _{01} & = & -ag\delta _{k_{y}-q_{y},0}\, \mathbf{M}_{k_{x}-q_{x}+\pi ,\omega -\nu}\cdot
 \vs _{\alpha \beta }, \\
\Sigma _{02} & = & \frac{a^{2}g}{2}\delta _{k_{y}-q_{y},\pi }\, 
\left(\mathbf{N}\left|
\mathbf{M}\right| ^{2}\right)_{k_{x}-q_{x}+\pi ,\omega -\nu }\cdot \vs _{\alpha \beta }, \\ 
\Sigma _{1} & = & -\frac{ag}{S}\mathbf{l}_{\mathbf{k}-\mathbf{q},\omega -\nu }\cdot\vs _{\alpha \beta },\\ 
\Sigma _{2} & = &
\frac{a^{2}g}{2S^{2}}\left(\mathbf{N}\left| \mathbf{l}
\right|^{2}\right)_{\mathbf{k}-\mathbf{q}+\mathbf{Q},\omega 
-\nu}\cdot \vs _{\alpha \beta },
\end{eqnarray}
 where $ \mathbf{Q}=\left( \pi/a ,\pi/a \right)  $ is the
antiferromagnetic modulation 
vector suitable for a ladder geometry. We also regroup the zero-th order term in
$ F^{-1}\equiv \hat{G}_{0}^{-1}-\Sigma _{00} $.

The evaluation of the various contribution in the continuum limit,
proceeds very similarly as in ref.~\cite{mu-zey}, and we refer to that 
paper for a more detailed explanation. The quantity to be evaluated is
%\begin{equation}
%S_{h\textrm{ eff}}=-\tr \ln \left( F^{-1}\right) -\tr \ln \left( \1
%-F\left( a\Sigma _{01}+a^{2}\Sigma _{02}+a\Sigma _{1}+a^{2}\Sigma _{2}\right) \right).
%\label{eq:trln}
%\end{equation}
\begin{equation}
S_{h\textrm{ eff}}=-\tr \ln \left( F^{-1}\right) -\tr \ln \left( \1
-F\left( \Sigma_{01}+\Sigma_{02}+\Sigma_{1}+\Sigma_{2}\right) \right).
\label{eq:trln}
\end{equation}
 We need then to find the inverse of $ F^{-1} $ up to $ O\left(
a\right)  $. It turns out that
\begin{equation}
F=\bar{F}D^{-1}-a\bar{F}D^{-1}RD^{-1}+O\left( a^{2}\right),
\end{equation}
where the various matrices are
\begin{eqnarray}
\bar{F} & = & \bar{g}^{-1}_{0}\left( \mathbf{k},\omega \right) \delta_{kq}
\delta_{\alpha \beta } -g\delta_{k_{y}-q_{y},\pi }\mathbf{N}_{k_{x}-q_{x}+\pi }\cdot \vs_{\alpha\beta } , \\
D & = & D (\mathbf{k},\omega) \delta_{kq} \delta_{\alpha \beta },\\ 
R & = & -g \, \delta_{k_{y}-q_{y},\pi } \sum_{r=x,\tau }
\left(k_r-q_r+\delta_{r,x}\pi/a \right) \, \partial_{r}g^{-1}_{0}( \mathbf{k},\omega )
\mathbf{N}_{k-q+Q}\cdot \vs _{\alpha \beta},
\end{eqnarray}
and we used the shorthand notation
\begin{eqnarray}
\bar{g}_{0}^{-1}\left(\mathbf{k},\omega_{n}\right) &=&g^{-1}_{0}\left(
\mathbf{k}+\mathbf{Q},\omega _{n}\right),\\
D(\mathbf{k},\omega_n )&=& g^{-1}_{0}\left( \mathbf{k},\omega_n \right)
\bar{g}_{0}^{-1}\left(\mathbf{k},\omega_{n}\right)-g^{2} .
\end{eqnarray}

We first consider the term 
\begin{equation}
\tr \ln \left( F^{-1}\right)=\tr \ln \left( \hat{G}^{-1}_0 \right)+\tr \ln
\left( \1 -\hat{G}_0 \Sigma_{00} \right). 
\end{equation}

The second term of this equation is reduced to the calculation of
\begin{equation}
\sum_{m=1}^\infty \frac{1}{n} \tr \left(\hat{G}_0 \Sigma_{00} \right)^m ,
\end{equation}
where each term has the following expansion
\begin{eqnarray}
\tr \left(\hat{G}_0 \Sigma_{00} \right)^m &=& (g)^m \sum_{k,q_2\ldots q_m}
g_0(k) \bar{g}_0(k+q_2)g_0(k+q_3) \bar{g}_0(k+q_4)\cdots g_0(k+q_{m-1})
\bar{g}_0(k+q_m)\nonumber\\
& & \times \mathbf{N}^{a_1}_{-q_2}\mathbf{N}^{a_2}_{q_1-q_2}\cdots
\mathbf{N}^{a_{2}}_{q_m}\tr \left(\sigma^a_1\sigma^a_2\cdots \sigma^a_m \right) ,
\label{trace-m}
\end{eqnarray}
with $m$ an even integer. The trace over the Pauli matrices can be
carried out using a trace reduction formula~\cite{pauli-trace}. The
gradient expansion in Eq.~(\ref{trace-m}) is then obtained by
performing an expansion of the product of propagators $g_0
(k)\cdots\bar{g}_0(k+q_m)$  in powers of the variables $q_2,q_3,\ldots q_m$ that appear as
argument of the vector field $\mathbf{N}$.
The result obtained is~\cite{mu-zey}

\begin{equation}
\tr \ln \left( \1 -\hat{G}_0 \Sigma_{00} \right)=
\int dxd\tau \left[ \frac{\bar{\chi}_{xx}}{2} 
\left| \partial_{x}\mathbf{N}\right|^{2}+\frac{\bar{\chi }_{\tau\tau}}{2}
\left| \partial_{\tau}\mathbf{N}\right|^{2}\right],
\end{equation}
with the definition
\begin{equation}
\bar{\chi }_{\alpha \beta }  =  \left. \frac{\partial ^{2}}{\partial q_{\alpha }\partial q_{\beta }}\sum
_{k}\ln \left[ 1-g^{2}g( k) g(
k+q+Q) \right] \delta_{q_{y}0}\right|_{q=0} .\label{chi0}
\end{equation}

We can now pass to the evaluation of the second term in
Eq.~(\ref{eq:trln}). This does not present particular problems, since
after expanding all the quantities, it reduces to the evaluation of a
finite number of traces. 
The result is
\begin{eqnarray}
\tr \ln \left( \1
-F\left(\Sigma_{01}+\Sigma_{02}+\Sigma_{1}+\Sigma_{2}\right) \right) 
& =& i\frac{g^{3}}{S}\int\!\!dx\,d\tau 
\hat{\chi}_\tau \left(\mathbf{N} \times \partial_\tau \mathbf{N}\right)\cdot 
\left(\mathbf{l}_1+\mathbf{l}_2\right) \nonumber \\
& & -\frac{g^{2}}{8S^{2}} 
\int\!\!dx\,d\tau 
\tilde{\chi}\left(\mathbf{l}_1+\mathbf{l}_2\right)^2.
\end{eqnarray}
Here we omitted to write a Gaussian term $ \propto \mathbf{M}^{2} $,
completely decoupled, which can be integrated out without further
consequences. The quantities $\hat{\chi}_\tau$ and
$\tilde{\chi}$ are given by
\begin{eqnarray}
\hat{\chi }_{\tau} & = & -i\sum _{k}D^{-1}(k)\partial_{\omega_n} 
g^{-1}_{0}(k)D^{-1}( k+Q) ,\label{chi1}\\ 
\tilde{\chi } & = & \sum _{k}\left[ D^{-1}(k)\left(
g^{-1}_{0}( k+Q) -g^{-1}_{0}(k)\right)
\right]^{2} . \label{chi2}
\end{eqnarray}

\begin{figure}
\begin{center}
\includegraphics[width=8cm,height=6cm]{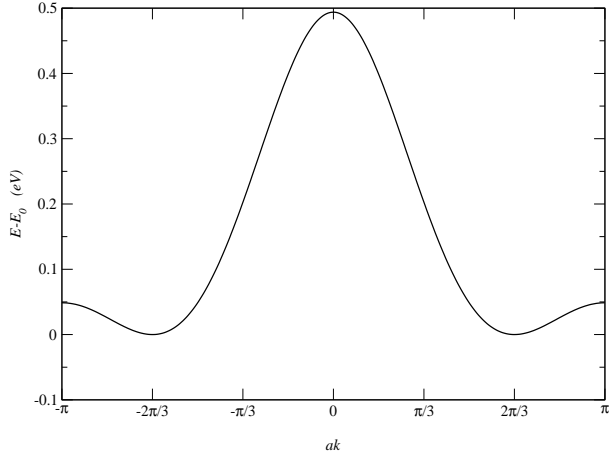}
\end{center}
\caption{\footnotesize Effective holes lowest-band emerging from our theory.
Parameters are $ t=0.24, J_K =1$ eV.The minimum falls exactly at $(ak)=2\pi/3$}
\label{MFband}
\end{figure}

They are generalized susceptibilities of
the holes in presence of long-wavelength spin fields. In particular the zeros
of $ D(k) $ determine the dispersion of such holes. The bands
originating in such a way correspond to free holes moving in 
a staggered magnetic field. Such a staggered field would
break translation invariance by one site and we would obtain four
bands in the reduced Brillouin zone.
Instead in our procedure we never broke
explicitly translation invariance, so that we obtain genuinely two bands in
the Brillouin zone. The lowest of these two band is symmetric in
character (bonding).  In Fig.\ \ref{MFband} we show it for
values of the constants relevant for the Copper-Oxide ladder \ie a
band-width of $\approx 0.5$ eV~\cite{\ARPES} and $J_K\approx
1$~\cite{mu-val-exp1,mu-val-exp2,mu-val-band}. This band is in  
good agreement with accurate calculations on the one hole spectrum of
the $t-J$ model. 
In particular, in the isotropic
$t-J$ model, for $t/J \approx 2$ the same qualitative feature are observed:
a global maximum at $(ka)=0$, global minima at $(ka)\approx\pm 2\pi /3$ and 
local maxima at $(ka)=\pm \pi$~\cite{oitmaa,miki}.

Now that we calculated the long wavelength contribution coming from
the holes, we still have to consider the continuum limit (in the low
energy sector) of the pure spin action $ S_{s} $ given by
eq. (\ref{eq:s-spin}). The result is   
\begin{eqnarray}
S_{s\textrm{ eff}} & = & -i\int\! dx\,d\tau \left( \mathbf{N}\times \partial _{\tau
}\mathbf{N}\right)\cdot \left(\mathbf{l}_1+\mathbf{l}_2 \right) 
+a\left(J_{\|} +\frac{J_\perp}{2}\right) \int\! dx\,d\tau 
\left(\mathbf{l}_1+ \mathbf{l}_2\right)^2 \nonumber\\ 
 &  & +aJ_{\|}\int\! dx\,d\tau \left(\mathbf{l}_1- \mathbf{l}_2\right)^2
+aS^2 J_{\|}\int dx\,d\tau \, \left|\partial_{x}\mathbf{N}\right|^{2} .
\end{eqnarray}

The very last step is the Gaussian integration of the 
$\mathbf{l}_{\perp } $ field, leaving us with the
effective long-wavelength action for the antiferromagnetic order parameter,
a (1+1) NL$ \sigma  $M:

\begin{equation}
S_{\textrm{eff}}=S_{h\textrm{ eff}}+S_{s\textrm{ eff}}=\frac{1}{2f}\int dxd\tau \left[
 v\left| \partial _{x}\mathbf{N}\right| ^{2}+\frac{1}{v}\left|
 \partial_{\tau }\mathbf{N}\right| ^{2}\right] ,
\end{equation} 
 where the NL$ \sigma  $M parameters are given by 
\begin{eqnarray}
f & = & \frac{1}{2}\left[ \left( S^2 J_{\|}-\frac{\bar{\chi
}_{xx}}{2}\right) \left( \frac{\left( 1+\frac{g^{3}}{S}\hat{\chi
}_{\tau }\right) ^{2}}{\left[4J_{||}+2J_\perp  
+\frac{g^{2}}{S^{2}}\tilde{\chi }\right]
}-\frac{\bar{\chi }_{\tau \tau }}{2}\right) \right] ^{-\frac{1}{2}}, \label{eqf}\\ 
v & = & a \left[ \frac{\left( S^2 J_{\|}-\frac{\bar{\chi
}_{xx}}{2}\right) }{\left( \frac{\left( 1+\frac{g^{3}}{S}\hat{\chi
}_{\tau }\right) ^{2}}{\left[4J_{||}+2J_\perp  
+\frac{g^{2}}{S^{2}}\tilde{\chi } \right]
}-\frac{\bar{\chi }_{\tau \tau }}{2}\right) }\right] ^{\frac{1}{2}} .\label{eqv} 
\end{eqnarray}

Hence, the spin-fermion model with mobile holes interacting with an
antiferromagnetic background is mapped
into an effective NL$ \sigma  $M whose coupling constant depend
on doping through the generalized susceptibilities in Eqs.\
(\ref{chi0}), (\ref{chi1}), and (\ref{chi2}).

Now we can immediately transpose to our model of a doped spin liquid,
some known result for the NL$ \sigma  $M, \eg 
mainly the presence of a gap which separates the singlet ground state
from a triplet of magnetic excitations. This gap should persist as
long as the continuum approximation is valid.

The fact that the NL$ \sigma  $M in (1+1) dimension has a gap above
the ground 
state can be established in a variety of ways. 
Using the two loop beta function~\cite{brezin76}
one obtains 
\begin{equation}
\Delta =v\Lambda e^{-\frac{2\pi }{f}}\left( \frac{2\pi }{f}+1\right),
\label{gap-eq} 
\end{equation}
where $\Lambda$ is a cutoff of the order of the inverse lattice constant.
Now we have an explicit analytic form for the doping dependence of the 
spin gap in the spin-liquid state of 
a two leg ladder. 

\begin{figure}
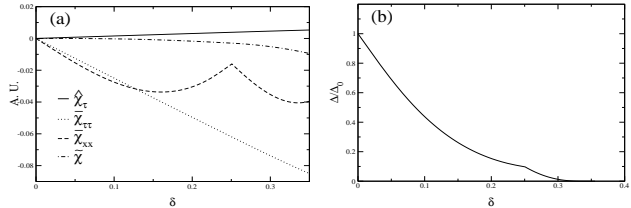

\subfigure{\includegraphics[width=4cm,height=2.7cm]{susc_gt.eps}}
\subfigure{
\includegraphics[width=4cm,height=2.7cm]{D_decs.eps}}
\caption{\footnotesize{$J_K > 2t$. 
(a) Generalized susceptibilities of Eqs.\ (\ref{chi0}), (\ref{chi1}), and
(\ref{chi2}) for $J_K=2,
\, t=0.76$ eV. For $J_K > 2t$ all the susceptibilities contribute to
lower $f$ hence the gap decreases for small doping for any value of
the constants.
(b) Normalized gap of eq.\ (\ref{gap-eq}). Here we fixed the exchange
constants to $J_{\parallel}=J_\perp=0.108$ eV. $\Delta_0$ is the gap
without  doping.  
\label{fig:decrease}}}
\end{figure}

\begin{figure}
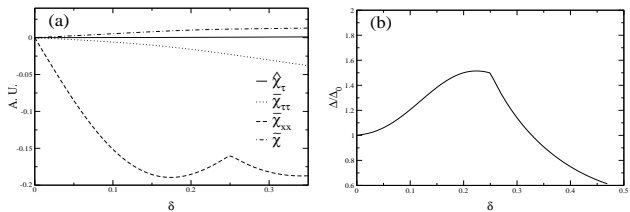

\subfigure{ 
\includegraphics[width=4cm,height=2.7cm]{susc_lt.eps}}
\subfigure{
\includegraphics[width=4cm,height=2.7cm]{D_grows.eps}}
\caption{\footnotesize{$J_K <2t$. 
(a) Generalized susceptibilities of Eqs.\ (\ref{chi0}), (\ref{chi1}), and
(\ref{chi2}) for $J_K=3 \,t=1.8$ eV. For $J_K<2t$, one susceptibility,
$\tilde{\chi}$, grows with doping  and contributes to increase $f$ and 
hence the gap. For $(J_K,t) \gg (J_{\parallel},J_\perp)$ we can have an
increasing gap for small doping.
(b) Normalized gap of eq.\ (\ref{gap-eq}). Fixing the exchange
constants to $J_{\parallel}=J_\perp=0.108$ eV is enough to have an
increasing gap for small doping.
\label{fig:increase}}}
\end{figure}

To study the behavior of the gap with doping we have to
distinguish two regimes where the lowest effective
band has minimum either at zero or at $2/3\pi$. For $J_K>2t$ the minima fall in
$\pm 2/3\pi$. Here all the generalized susceptibilities in Eqs.\
(\ref{chi0}), (\ref{chi1}), and (\ref{chi2}) contribute to lower $f$
and, since from eq.\ (\ref{gap-eq}) $\Delta$ is an increasing function
of $f$, they make the gap smaller for any value of the
constants (see Fig.~\ref{fig:decrease}). This is comforting, since,
as we mentioned, for $J_K$ very  
large the physics of the Spin-Fermion model should be similar to that
of the $t-J$ model~\cite{ZR-pd-tj}, and for that one, TDMRG simulations 
show that the gap decreases at least in a strong anisotropic case
($J_\perp =10 J_{\|}$). 
When $J_K<2t$ the band minimum falls in zero and there is
one susceptibility, $\tilde{\chi}$, which instead makes $f$ grow.
$f$. In this regime there is then a (small) region of parameters where 
the gap grows with doping (see figure~\ref{fig:increase}).

Before passing to a comparison
with experiments, we want to comment on a possible simplifying
understanding. 

A simple picture to explain the observed diminishing of the spin gap
with doping in
$\textrm{Sr}_{14-x}\textrm{Ca}_{x}\textrm{Cu}_{24}\textrm{O}_{41}$, is 
that (at least for low doping concentration where speaking of a spin
liquid is still feasible) the effect of the holes is that of
renormalizing the anisotropy parameter $\lambda=J_\perp/J_{\|}$ for the spin part 
towards larger
values. In many studies on the 2 leg ladder Heisenberg
antiferromagnet~\cite{gopalan,reigrotzki,greven}, the spin gap is seen to
increase with $\lambda$. In fact, the same occurs in the \NLSM \ 
without doping in the range $\lambda \approx 1\div 2$. 

We can now pass to our mapping of a doping spin liquid to an
effective NL$\sigma$M.
According to equations (\ref{eqf},\ref{eqv}) effective coupling constants
$\tilde{J}_{\|}, \tilde{J}_\perp$   can be defined for the doped system
such that the form of the NL$\sigma $M parameters is that for a pure spin
system~\cite{morandi} \ie
\begin{eqnarray}
f&=&\frac{1}{S} \sqrt{1+\frac{\tilde{J}_\perp}{2\tilde{J}_{\|}}}, \\
v&=&2aS\tilde{J}_{\|} \sqrt{1+\frac{\tilde{J}_\perp}{2\tilde{J}_{\|}}}.
\end{eqnarray}

A small doping expansion in the regime $J_K>2t$ leads to 
\begin{eqnarray}
\tilde{J}_{||} & = & J_{||}+\frac{3}{4}\frac{\left(
J^{2}_{K}-4t^{2}\right) }{J_{K}}\delta +O\left( \delta ^{2}\right), \\ 
\tilde{J}_{\perp } & = & J_{\perp }-\left( \frac{3}{2}\frac{\left(
J^{2}_{K}-4t^{2}\right) }{J_{K}}+2\left( 4J_{||}+2J_{\perp }\right)
+\frac{\left( 4J_{||}+2J_{\perp }\right) ^{2}}{8J_{K}}\right) \delta +O\left( \delta
^{2}\right) ,  
\end{eqnarray}
so indeed $\tilde{J}_{\|}$, $\tilde{J}_\perp$ are seen respectively to
increase, decrease, such that $\lambda$ decreases. However, such an
interpretation breaks down beyond $\delta\approx 0.04$ whereas $f,v$ are still
well defined positive constants.
This means that beyond such doping, this simplified picture
cannot be na{\"\i}vely applied and holes have a more effective way of
lowering the gap.

\section{Comparison with Experiments}

We come now to the comparison with experiments. Our theory depends on
four parameters $t, J_{K}, J_{\parallel}, J_{\perp}$ which we now want
to fix to physical values. ARPES experiment on $
\textrm{Sr}_{14}\textrm{Cu}_{24}\textrm{O}_{41} $ were performed
by Takahashi \etal~\cite{takahashi} who found  a band matching the
periodicity of the ladder with a bandwidth 
of  $\sim 0.5 \div 0.4$ eV. Adjusting our lowest band to have such a
bandwidth we obtain a relation between $t$ and $J_{K}$. On the other
hand, experiments on the CuO$_2$ cell materials and band theory
calculation~\cite{mu-val-exp1,mu-val-exp2,mu-val-band} agreed in 
assuming a value of $J_K$ of the order of $J_K\approx 1\div2$ eV. This in turn
gives us a value of $t\approx0.24\div 0.76$ eV, which is also consistent with
the same calculation.

\begin{figure}
\begin{center}
\includegraphics[width=8cm,height=6cm]{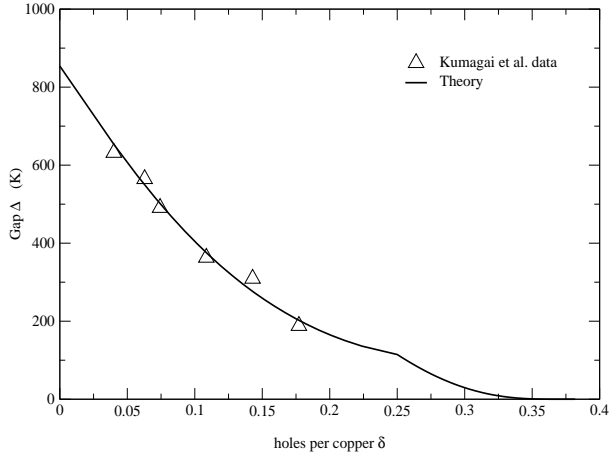}
\end{center}
\caption{\footnotesize Result of our theory and comparison with
experiments. The values of the constants used in equation (\ref{gap-eq})
are $ t=0.76, J_K =2$ eV, $J_\parallel /J_\perp=1$. The momentum cutoff
$\Lambda$ was 
fixed by fixing the the value of the gap with the one measured in 
$\textrm{Sr}_{14}\textrm{Cu}_{24}\textrm{O}_{41}$. For anisotropic
case $J_\perp/J_\parallel=0.8$  
the curve does not change appreciably.}
\label{fig-theo_kumagai}
\end{figure}

The debate around an anisotropy of the spin 
exchange constants~\cite{\INS} in $
\textrm{Sr}_{14-x}\textrm{Ca}_{x}\textrm{Cu}_{24}\textrm{O}_{41} $
seems now to be resolved in favor of isotropy or light anisotropy of
the coupling constant: $J_{\perp}/J_{||}\approx 0.8$~\cite{gozar}. 
We adjusted the value of the momentum cutoff $\Lambda$ by fixing the
theoretical gap with the 
experimental one for the undoped compound
$\textrm{Sr}_{14}\textrm{Cu}_{24}\textrm{O}_{41}$. 
Finally, to compare with the measured values of the gap for different
doping concentration $x$ in
$\textrm{Sr}_{14-x}\textrm{A}_{x}\textrm{Cu}_{24}\textrm{O}_{41}$
(where $A$ can be either divalent $\textrm{Ca}^{2+},\textrm{Ba}^{2+}$ 
or trivalent $ \textrm{Y}^{3+}, \textrm{La}^{3+}$), we still need a
relation between the A substitution $x$ and the number of holes
per copper site present in the ladder $\delta$.
This is another unsettled issue of the telephone number compound. In
particular Osafune \etal~\cite{osafune97} studying the optical
conductivity spectrum, inferred that with increasing Ca substitution $x$, 
holes are transferred from the chain to the ladder. On the other hand
N{\"u}cker \etal~\cite{nuecker00} argue that in the series compound
$\textrm{Sr}_{14-x}\textrm{Ca}_{x}\textrm{Cu}_{24}\textrm{O}_{41}$ 
the number of holes in the ladder is almost insensitive to Ca
substitution $x$ (although a small increase is observed).
Here we will assume that
$\textrm{Sr}_{14-x}\textrm{A}_{x}\textrm{Cu}_{24}\textrm{O}_{41}$  is
an example of doped spin  liquid and will use the data
from~\cite{osafune97}. The result
of our theory can be seen in figure~\ref{fig-theo_kumagai}.
There we used isotropic exchange constant, but the theoretical curve
did not change in a visible way if an anisotropy of $J_{\perp}/J_{||}\approx
0.8$ was inserted. 
We see from the figure that the spin gap becomes zero for $\delta\approx0.37$,
beyond this value the coupling constants $f$ and $v$ would become
imaginary signaling that our effective model cease to make sense. This 
means that for such doping ratios our parameterization~(\ref{eq:param1}) is 
no longer valid, in the sense that it does not incorporate the most
important spin configurations. However our theory could cease to make
sense much 
before. If one takes the point of view of the $t-J$ model (as we said the
Spin-Fermion model should map to it for large $J_K$) the holes
introduced in the system couple rigidly to the spins forming singlet
with the $P_i$ states. In the worst case this would limit the
correlation length of the spin to the mean hole-hole distance
$1/\delta$. In our case this happens at a doping ratio of $\delta\approx 0.15$.

A word of caution should be mentioned with respect to comparison with
experimental results. A still unresolved controversy is present
between NMR~\cite{\NMR} and neutron scattering~\cite{\INS}
experiments, where the latter see 
essentially no doping dependence of the spin gap. Without being able
to resolve this issue, we would like, however, to stress, that beyond
the uncertainties in experiments, the doping behavior obtained for the
spin-gap agrees with the numerical results in TDMRG and is opposite to 
the one obtained in mean-field treatments, making clear the relevance
of fluctuations.

\section{Conclusion}

In this paper we studied the behavior of the spin gap of a two leg
Heisenberg antiferromagnetic ladder as microscopically many holes are
introduced in the system. 
Such a situation can be physically realized in the series compound 
$\textrm{Sr}_{14-x}\textrm{A}_{x}\textrm{Cu}_{24}\textrm{O}_{41}$ with 
A=Ca, Y, La, and numerous result are now available from
experiments. On the theoretical side, however,
there is a contradiction between 
previous analytical treatments on the one hand, and TDMRG
simulations or NMR experiments on the other hand. Whereas in the first 
case, a magnon gap increasing with doping is predicted, a decrease is
observed in accurate numerical simulation and experiments.

Starting from the spin-fermion model we were able to solve the
contradiction using a controlled analytical treatment that properly
takes into account fluctuations in the continuum limit. Integrating
out the fermions we were left with a Fermi-determinant which we can
evaluate exactly in that limit.
The result is a non linear $\sigma$ model with doping dependent parameters.
The spontaneously generated mass gap
of this theory is seen to decrease as holes are introduced. Once
physical value for the parameters are given, we obtained very good
agreement with NMR experiments performed on Sr$_{14-x}$A$_x$Cu$_{24}$O$_{41}$.

\begin{acknowledgments}
Support by the Deutsche Forschungsgemeinschaft under Project No.\ Mu~820/10-2 
is aknowledged.
\end{acknowledgments}

%\bibliography{two_leg}

\end{document}